\documentclass{article}

\usepackage{arxiv}

\usepackage[utf8]{inputenc} 
\usepackage[T1]{fontenc}    
\usepackage{hyperref}       
\usepackage{nicefrac}       
\usepackage{microtype}      
\usepackage{lipsum}
\usepackage{graphicx}
\usepackage{subfigure} 
\usepackage{color}
\usepackage[english]{babel} 
\usepackage{amsmath,amsfonts,amsthm} 
\usepackage{float}
\usepackage{algorithmic}
\usepackage{enumerate} 
\usepackage{array}
\usepackage{multirow}
\usepackage{xfrac}
\usepackage[table]{xcolor}
\usepackage{lineno}
\definecolor{lg}{gray}{0.9}

\usepackage{url}

\title{Temporal convolutional networks and transformers for classifying the sleep stage in awake or asleep using pulse oximetry signals}

\author{
  Ramiro Casal \\
  Lab. Señales y Dinánicas no Lineales\\
  IBB - UNER - CONICET \\
  Oro Verde (3100), Entre Ríos, Argentina  \\
  \texttt{rcasal@conicet.gov.ar} \\
  \And
  Leandro E. Di Persia \\
  sinc(i)\\
  FICH - UNL - CONICET \\
  Santa Fe (3000), Santa Fe, Argentina \\
  \texttt{ldipersia@sinc.unl.edu.ar} \\
  \AND 
  Gastón Schlotthauer \\
  Lab. Señales y Dinánicas no Lineales\\
  IBB - UNER - CONICET \\
  Oro Verde (3100), Entre Ríos, Argentina  \\
  \texttt{gschlotthauer@ingenieria.uner.edu.ar} \\
}

\begin{document}
\maketitle

\begin{abstract}
Sleep disorders are very widespread in the world population and suffer from a generalized underdiagnosis, given the complexity of their diagnostic methods. Therefore, there is an increasing interest in developing simpler screening methods. A pulse oximeter is an ideal device for sleep disorder screenings since it is a portable, low-cost and accessible technology. This device can provide an estimation of the heart rate (HR), which can be useful to obtain information regarding the sleep stage. In this work, we developed a network architecture with the aim of classifying the sleep stage in awake or asleep using only HR signals from a pulse oximeter. The proposed architecture has two fundamental parts. The first part has the objective of obtaining a representation of the HR by using temporal convolutional networks. Then, the obtained representation is used to feed the second part, which is based on transformers, a model built solely with attention mechanisms. Transformers are able to model the sequence, learning the transition rules between sleep stages. The performance of the proposed method was evaluated on Sleep Heart Health Study dataset, composed of $5000$ heathy and pathological subjects. The dataset was split into three subsets: $2500$ for training, $1250$ for validating, and $1250$ for testing. The overall accuracy, specificity, sensibility, and Cohen's Kappa coefficient were $90.0\%$, $94.9\%$, $78.1\%$, and $0.73$.\end{abstract}

\keywords{Automatic sleep staging \and Pulse oximetry \and Attention models \and Transformers}

\section{Introduction} \label{intro}

Sleep plays a fundamental role in humans, strongly influencing numerous vital processes related to the regulation of the function of the immune and endocrine systems, the restoration of the musculoskeletal system, the consolidation mechanisms of memory, the preservation of cognitive function, amongst many others \cite{stickgold2005sleep}. Sleep disorders include more than $80$ different pathologies that affect a significant portion of the world adult population, for instance, in US approximately $35-40\%$ of adult people report having difficulty falling asleep. The high prevalence of sleep disorders and their consequent health implications and economic costs are the principal reasons for the constant growing interest in this subject, not only in clinical practice but also in research area \cite{sateia2014international, skaer2010economic}. 

Sleep and its pathologies are evaluated in sleep medical centers through polysomnography (PSG), a study where several physiological signals are recorded simultaneously while the patient is asleep. Then, these recordings are manually scored, sometimes with the help of a software, in order to determine the severity of varied pathologies. The scoring is performed according to two available standards: the traditional Rechtschaffen and Kales (R\&K) \cite{rechtschaffen1968manual} and, since 2007, the American Academy of Sleep Medicine (AASM) \cite{berry2012aasm}. Both manuals represent a guideline to score several events --namely arousal, muscular movements, cardiac and respiratory events, sleep staging-- that allow to characterize sleep. The R\&K standard define the following sleep stages: \emph{wakefulness} (W), two stages of \emph{light sleep} (N1 and N2), two of \emph{deep sleep} (N3 and N4) and \emph{rapid eye movement sleep} (REM). More recently, the AASM unifies N3 and N4 in a single stage, called simply N3 or slow wave sleep, with the intention of reducing the inter-professional discrepancies. The PSG recordings are split in $30$ seconds long segments and classified into the sleep stages on the basis of characteristic waveforms that can be found in electroencephalography (EEG), electromiography (EMG) and electrooculography (EOG). 

Many efforts have been made in the scientific community to automate sleep staging with the aim of overcoming the disadvantages of tedious manual scoring, that is, in addition, susceptible to human errors, showing high inter-professional variability \cite{norman2000interobserver}. Automatic sleep staging is performed primarily using the EEG signal, but sometimes information is also extracted from other signals such as EOG and EMG \cite{supratak2017deepsleepnet, phan2019seqsleepnet, fraiwan2012automated}. Nevertheless, there is an increasing demand on home-based devices to obtain sleep stages classification, without using EEG. This can be useful for broad screenings that allow evaluating the prevalence of sleep disorders in large population, avoiding the overload of the health systems and, consequently, the long waiting lists and the under-diagnosis that are commonplace in these diseases \cite{fuhrman2012symptoms}.

The most widespread proposals for automatic screening of sleep staging use cardiac signals \cite{penzel2003dynamics, aeschbacher2016heart}. Various studies have found that average heart rate ($\text{HR}$) is highest during wakefulness, decreasing steadily from light to deep stages and then increasing again in REM sleep. Furthermore, $\text{HR}$ variability is higher during W and REM stages than N1, N2 and N3 \cite{penzel2003dynamics}. These findings established the grounds to develop algorithms to classify sleep stages using cardiac-related signals, formerly ECG \cite{adnane2012sleep, yucelbacs2018automatic, malik2018sleep}. However,  more recent papers have been published using photoplethysmography (PPG) from pulse oximeter instead, with the argument of being a more portable, simpler and low-cost technology than ECG \cite{uccar2016automatic, casal2019hely, beattie2017estimation}. In these screening works, sleep stages are usually grouped in order to achieve a primary result that allows to estimate patient's condition, to be then transferred to a complete study if necessary. The most common is to classify in W, REM and non-REM stages (grouping N1, N2 and N3) or directly in W and asleep (S) (grouping REM and non-REM stages). A detailed comparison of all these works will be addressed in the next section. 

This work is the third part of a series of previously published articles that address the same problem \cite{casal2019hely,casal2019rnn}. In our first work \cite{casal2019hely}, we extracted hand-crafted features from $\text{HR}$ signals obtained with a  pulse oximeter. These features were designed in order to recognize the characteristic patterns corresponding to W and S stages, using entropy and complexity measures, frequency domain and time-scale domain techniques and classical statistics. The features are then used to train different classifiers for determining the sleep stage.

With the emergence of deep learning, conventional hand-crafted features approaches have been gradually replaced by automatic-learned features from the input data. The most popular architecture for extracting features automatically is convolutional neural networks (CNN), in which learned filters are used to convolve with the raw input data in order to obtain time-invariant features. Multiple layers, principally composed of these filters, non-linear functions and pooling operations, are applied sequentially to learn representations of the data with multiple levels of abstraction \cite{lecun2015deep}. Nevertheless, for tasks that involve sequential inputs, like automatic sleep staging, recurrent neural networks (RNN) were the default option since they are able to store information about the history of all the past elements of the sequence in the hidden units, learning the temporal dependencies \cite{lecun2015deep}. This aspect was kept in mind in our second work \cite{casal2019rnn}, in which we applied a RNN to $\text{HR}$ signals obtained with a pulse oximeter to classify the sleep stage. The proposed architecture consisted of two stacked bidirectional gated recurrent units (GRUs) and a softmax layer to classify the output. In spite of outperforming the previous result, this work has several disadvantages associated with RNN. The sequential nature of RNN impedes parallelization, this leads to a slow training process, generating large memory constraints \cite{vaswani2017attention}. Furthermore, the training process is highly unstable.

In this paper we designed a network architecture based on attention mechanisms to overcome the RNNs disadvantages, achieving a faster training without losing performance. Attention networks allow modeling temporal dependencies independently to the distance in the inputs or outputs sequences, being more hardware-friendly than RNNs \cite{vaswani2017attention}. Since most of these architectures have been designed for natural language processing, the inputs are usually high-dimensional due to the token embedding. As this is not the case of pulse oximeter signal (1D-signal if only $\text{HR}$ from pulse oximeter is considered), we used a temporal convolutional network --a generic architecture that combines CNN and residual neural networks-- to map the input sequence to a higher dimension while reducing the length sequence \cite{bai2018empirical}. In this way, the input data is able to be processed by the attention network using a sequence whose length is shorter than the raw input and, consequently, faster to learn.

The aim of the current work is to classify the sleep stage in W and S using pulse oximeter signals, achieving a reliable and fast outcome. The main contributions of this works are the following:

\begin{itemize}
 \item We present an architecture based on TCN and attention mechanisms to perform a sequence-to-sequence classification, neither using RNN or hand-engineer features nor decreasing performance compared to the state of the art.
 \item We use a reduced dataset based only on $\text{HR}$ obtained from pulse oximeter to perform a simplified sleep stage classification, achieving a comparable result with works that use more informative signals. These results permit a fast screening of sleep disorders.
 \item We compare three different methods for processing sequential inputs applied to sleep staging: a classical approach using hand-engineered features and classifiers, an RNN-based proposal and, finally, a method built with TCN and attention mechanisms. 
\end{itemize}


\section{Proposed architecture} \label{sec:prop_arch} 

The architecture of the designed network comprises two fundamental parts: the former for representation learning and the latter for sequence modeling. The first part --based principally on TCN-- was designed to obtain a representation of the $\text{HR}$ signals to be used as input in the second part. This second part --based mainly on transformer, a model which is solely built with attention mechanisms-- was developed to learn the transition rules between sleep stages in the sequence. This whole architecture receives as input the raw $\text{HR}$ signal and produces as output the classification in W and S every 30-seconds segments.

\subsection{Representation learning}

TCNs are a family of architectures based on simple CNNs, which have emerged with the aim of comparing the performance of RNN and CNN for sequence modeling tasks, assessing the use of the former as a default starting point. The original TCN proposal by Bai et al. \cite{bai2018empirical} uses a 1D fully-convolutional network, where each layer maps the input to a same-length output, achieving this with zero-padding. The considered convolutions are causal, meaning that the outputs for a given time only depend on previous inputs. Furthermore, the authors employed dilated convolutions that enable an exponentially large receptive field. The receptive field depends on the dilated convolutions, the filter size in the CNN and the depth of the network \cite{bai2018empirical}.

TCN is built with basic blocks --called residual blocks--, which consist of two layers of dilated convolutions with a rectified linear unit (ReLU) as non-linear function \cite{nair2010rectified}. Each residual block has a residual connection \cite{he2016deep}, which allows to learn only the modification in the identity mapping (i.e. the residual) rather than the complete output, simplifying the convergence in very deep networks. 

In this work, we used a TCN-inspired architecture to automatically extract features from the $\text{HR}$ signals. This part of the network is designed with two different aims. The first aim is to learn a high-dimensional representation of the input with which to feed the forthcoming part of the network. Along with this representation learning, as we have already stated, the second aim is diminish the sequence length by dimensionality reduction over time, for the sake of computational cost.

We introduced some modifications to the TCN in order to achieve these objectives. Firstly, we did not considered causal convolutions since online signal processing is not necessary. With reference to this, we used bidirectional GRUs to be able to exploit both past and future information in our previous paper \cite{casal2019rnn}. Secondly, we included in the architecture a stride operation --recent works suggests that pooling can be replaced by stride without a performance degradation \cite{he2016deep, springenberg2014striving}-- along with the convolution in pursuit of the mentioned second aim. Lastly, for normalization and regularization, we applied batch normalization \cite{ioffe2015batch} and dropout \cite{hinton2012improving}.

\begin{figure}[!t]
\centering
\includegraphics[width=0.5\textwidth]{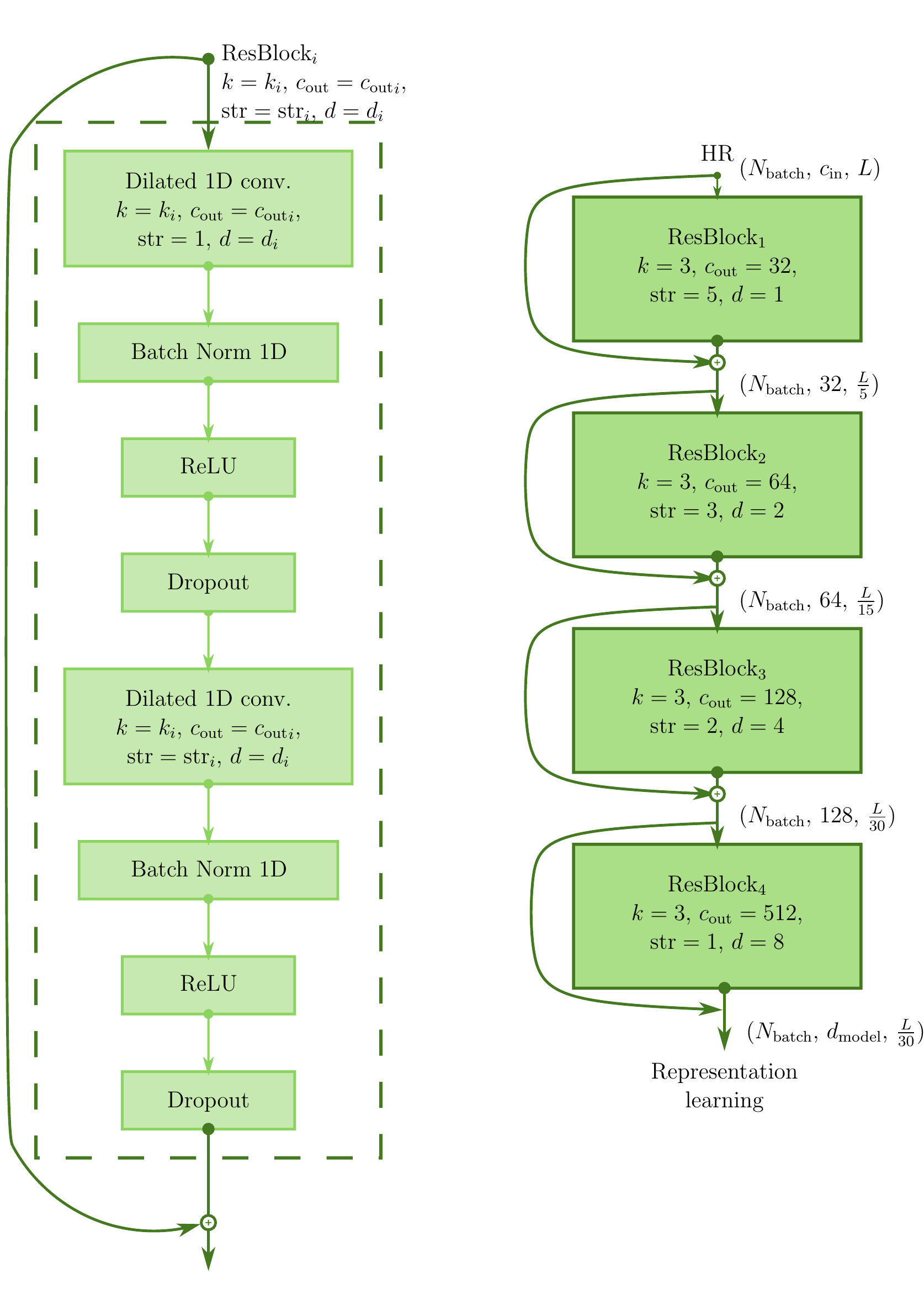}
\caption{An overview architecture of the representation learning part in the network. On the left is the residual block composition and its parameters. On the right is the designed network which consist of four residual blocks. After each layer, the dimension transformation performed is shown.}
\label{fig1}
\end{figure}

The layout of this part of the network is shown in the Fig. \ref{fig1}. In our model, each residual block consists of two convolutional sublayers, which performs four operations sequencially: dilated 1D convolution --with filters of size $k$, number of output channels (or number of filters) $c$, dilation parameter $d$ and stride $\mathrm{str}$--, batch normalization, ReLU activation and dropout operation. Dimensionality reduction is only performed on the second sublayer using the stride, while in the first convolutional layer stride is set equal to $1$. 

This part of the network consist of four residual blocks, which parameters can be found in Fig. \ref{fig1}. As it can be seen, each residual block increases the number of channels gradually from the 1D dimension to $d_\mathrm{model}$ as well as reduces the sequence length through strides from $L$ to $L/30$. Thus, time-invariant and high-dimensional features are the output of the fourth residual block, extracted from the raw $\text{HR}$ signal. We called these features ``representation learning''.

\subsection{Sequence modeling}
For the second part of the network, we applied an encoder-decoder structure to perform the sequence modeling based on a type of architectures called tranformer, which only uses attention mechanisms \cite{vaswani2017attention}. The encoder maps an input sequence to a representation sequence, which is then used to feed the decoder for generating the output sequence one element at a time. In our model, the encoder was used as in Vaswani et al. \cite{vaswani2017attention}, but the decoder was simplified to a fully connected layer for interpretation and classification of the encoding sequence. 

Most of the transduction models have been designed for natural language processing and because of this they use learned embedding to convert words to vectors of dimension $d_{\mathrm{model}}$. To adapt this transformation to our problem, we used the TCN architecture described above for obtaining a representation learning of dimension $d_{\mathrm{model}}$ --first aim in representation learning part--. Next we will describe all the elements of this sequence modeling part following the same organization used by Vaswani et al. \cite{vaswani2017attention}. Also, it is shown in the Fig. \ref{fig2}.

\subsubsection{Encoder}

The encoder is composed by two stacked layers, each one containing two sublayers. The first sublayer consist of a multi-head attention mechanism and the second is a feed-forward network (FNN). Both sublayers have a residual connection followed by a layer normalization \cite{ba2016layer}.

To calculate self-attention, the encoder inputs are packed together in a matrix which is projected by three previously trained matrices $\mathbf{W}^Q$, $\mathbf{W}^K$, $\mathbf{W}^V$. As a result, three matrices are obtained, which are called query $\mathbf{Q}$, key $\mathbf{K}$ and value $\mathbf{V}$. Each row of these matrices correspond to the projection of each input element. Afterward, the attention matrix is computed as follow:

\begin{equation}\label{eq:attention}
 \mathrm{A}(\mathbf{Q},\mathbf{K},\mathbf{V}) = \mathrm{softmax}(\frac{\mathbf{Q} \mathbf{K}^T}{\sqrt{d_k}})\mathbf{V}
\end{equation}

\noindent where $\sqrt{d_k}$ is the dimension of $\mathbf{K}$ and $\frac{1}{\sqrt{d_k}}$ is a scaling factor that pushes the dot product to a sensitive region of the softmax function. 

We employed multiple attention mechanisms as in \cite{vaswani2017attention}, using $h$ different randomly initialized matrices $\mathbf{W}^Q_i$, $\mathbf{W}^K_i$, $\mathbf{W}^V_i$ to obtain $h$ input projections. This modification allows the model to have multiple representation subspaces from which to calculate $h$ attention matrices. Then, these $h$ attention matrices are concatenated and multiplied by another learned matrix $\mathbf{W}^0$ to produce the output of the multi-head attention block, which is finally used to feed the second sublayer --that is, the FNN--. 

In this work we employed $h=8$, reducing the dimension ${d_k}=d_{\mathrm{model}}/h$ in order to keep relatively constant the computational cost.

The FNN in the second sublayer is composed by two linear transformation with a ReLU activation, calculated as $\mathbf{W}_2 \textrm{ReLU}(\mathbf{W}_1 x + b_1) + b_2$. The dimensionality of inputs and outputs is $d_{\mathrm{model}}$ and the dimensionality of the inner-layer is $d_{\mathrm{ffn}}=512$.

Positional encoding is the last element that we have to address for understanding the encoder architecture. We calculated positional encoding as follow \cite{vaswani2017attention}:

\begin{equation}
\begin{aligned} \label{Eq:pos_encoding}
 \mathrm{PE}(p, 2i)   &= \sin (p/10000^{2i/d_model}),  \\ 
 \mathrm{PE}(p, 2i+1) &= \cos (p/10000^{2i/d_model})
\end{aligned}
\end{equation}

\noindent where $p= [0,\, 2, \, \ldots, \, L_{\max}-1]$ is the element position in the sequence and $i=[0,\, 2, \, \ldots, \, d_{\textrm{model}}-1]$ is the embedding index.  $L_{\max}$ is the length of the longest sequence. The positional encoder allows to the model to learn a specific pattern which contains information about the relative position of the $p$-th element in the whole sequence, representing the order of the sequence. To be added, both the encoder input and the positional encoding have the same dimension.

\subsubsection{Decoder}
The encoder was designed in order to learn the transition rules in the sequence. This information is then classified by the decoder. The decoder simply consists of a FNN composed by two linear transformation with a ReLU activation in between, calculated as $\mathbf{W}_2 \textrm{ReLU}(\mathbf{W}_1 x + b_1) + b_2$. The input and output dimension of the first layer is $d_{\mathrm{model}}$. In the second layer, the output dimension is reduced to the number of considered sleep stages, that is $2$. Finally, a softmax layer was used for classification.

 \begin{figure}[!t]
\centering
\includegraphics[width=0.5\textwidth]{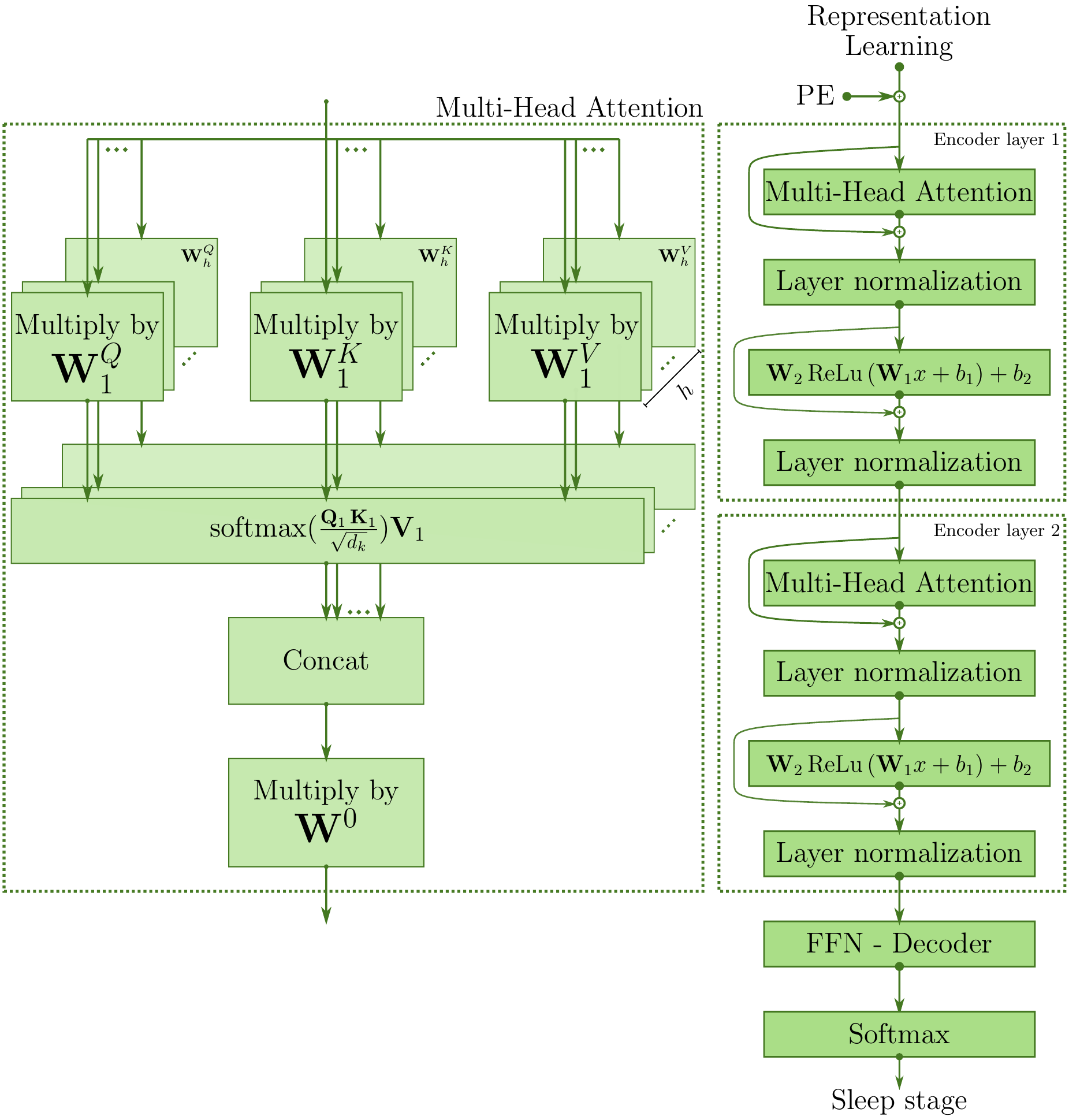}
\caption{An overview architecture of the sequence modeling part in the network. On the left is the multi-head attention block. On the right is the described network which consist of two stacked encoder layers and a decoder layer.}
\label{fig2}
\end{figure}

\section{Experiments} \label{sec:experiment}

\subsection{Database description}

In our work, we evaluated the designed model using $\text{HR}$ signals from the \emph{Sleep Health Heart Study} database (SHHS). These signals belong to polysomnography studies acquired by home-based devices with supervision from specialized technicians \cite{redline1998methods}. 

SHHS dataset belongs to a study designed to determine cardiovascular and other consequences of sleep-disordered breathing, testing its association with risk of coronary hearth disease, strokes and hypertension. More than $5000$ home-PSG registers were obtained from participants of the study who were included in the admission criteria. This first set of PSG --called SHHS 1-- was used as baseline to be compared with a follow-up set of PSG --SHHS 2--, which was obtained several years later recording the same subjects. SHHS 2 only contains $4080$ registers, as not all subjects repeated the study. 

The SHHS recordings were processed according with manuals of sleep scoring and associated events \cite{rechtschaffen1968manual, berry2012aasm}, providing a first output through software that was later corrected by hand. A comprehensive description about the SHHS database is in \cite{redline1998methods, nieto1997sleep}. 

In this work, we only used SHHS 1, discarding SHHS 2 to avoid the presence of repeated subjects in the data. We generated three subsets by randomly selecting $5000$ subjects belonging to SHHS 1 dataset. The first subset consist of $2500$ subjects for network training. The second and third subsets contains $1250$ subjects each for validation and testing of the trained network. In Table \ref{tab:shhs1} we summarized the principal characteristics of the SHHS 1 dataset.  

\begin{table}[t]
\centering
\caption{Characteristics of the study population in SHHS 1.}
\begin{tabular}{l l }
\hline
                                & SHHS 1 (min, max)  \\
\hline
Total subjects                  & $5080$ \\
Age                             & $63.1 \pm 11.2  \quad (39.0,\, 90.0)$ \\
Percentage of woman             & $52.3\%$ \\
Epworth sleepiness scale        & $7.8 \pm 4.4    \quad (0.0,\,  24.0)$ \\
Arousal index(/hr)              & $19.2 \pm 10.7  \quad (0.0,\, 110.4)$ \\
Apnea/hypopnea Index (/hr)      & $9.6  \pm 12.7  \quad (0.0,\, 115.8)$ \\
Total sleep time  (min)         & $587.7 \pm 107.6 \quad (35.0,\, 858.0)$ \\
Time awake / Time sleeping           & $0.742$ \\
\hline 
\end{tabular}
\label{tab:shhs1}
\end{table}

The $\text{HR}$ signal used in this work was acquired using a pulse oximeter Nonin XPOD Model 3011, which has a sample rate of $1$ Hz and a precision of $\pm 3$ beats per minute. The oximeter also provides an additional quality status signal, the value of which is zero for good connection and two for defective connection.

\subsection{Implementation details}

In this subsection the experiment setup and training parameters will be described.

\subsubsection{Hardware and software frameworks}

We implemented the model proposed in this paper using Python 3 and the an open-source machine learning library Pytorch.. The experiments were run on multiple Graphics Processing Units (GPUs) installed on a high-performance cluster and a personal computer, which are equipped with three GPU NVIDIA GTX 1080Ti and one GPU NVIDIA Titan Xp, respectively.

\subsubsection{Preprocessing and batching data}

As we have already stated, SHHS 1 dataset was splitted in three independently subsets for training, validation and testing, each of them with $2500$, $1250$ and $1250$ different subjects, respectively. The whole signals were preprocessed before being used in the proposed architecture. Firstly, we identified the invalid-HR segments, using the complementary quality status signal provided by the oximeter. Once identified, the unreliable data was replaced by linear interpolation between the previous and posterior confident HR estimations. Secondly, we standardized the HR signals to have zero mean and unit variance. For this purpose, we extracted the global mean and standard deviation from the training dataset. Then, these values were used for standardize the three subsets for training, validation and testing.  

It is common in most deep learning frameworks to group the input data in mini-batches, which are used to calculate model error and update model coefficients by the optimizer. Thus, the input data grouped in a batch was zero-padded to have the same length that the longest sequence of the batch. In order to avoid an excessive padding, the inputs were previously sorted by length. On this wise, data that belong to a same batch have similar lengths. For the batch size, we varied $N_{\textrm{batch}}$ from $2$ to $32$. The best performance was obtained for the value $16$.

\subsubsection{Optimizer and regularization}

The model was trained using Adam optimizer, an extension to stochastic gradient descent \cite{kingma2014adam}. The parameters in this optimizer are learning rate $\alpha$ and the coefficients used for computing the exponential decay rate for the first and second moment estimates, $\beta_1$ and $\beta_2$. We varied $\alpha$ from $10^{-2}$ to $10^{-6}$, obtaining the best performance for the value $10^{-4}$. For $\beta_1$ and $\beta_2$, we tried with several values near to the default values reported in the literature. We set these values equal to $0.9$ and $0.99$, respectively. 

Moreover, we used the L2 regularization, which modifies the loss function by adding a penalty term to avoid an increase in the magnitude of the model coefficients. The weight decay parameter is the coefficient which multiplies the L2 term, and was set equal to $10^{-3}$, which was obtained after varied the value from $10^{-3}$ to $10^{-5}$. 

Finally, we used dropout, which is a regularization technique that randomly sets individual nodes of the network to zero during training with $p$ probability \cite{srivastava2014dropout}. Thus, the network is forced to learn more robust features, avoiding over-fitting. We varied the probability $p$ from $0$ to $0.9$ in steps of $0.1$. The best performance was obtained with $0.1$ and $0.8$. The training was more stable with the value $0.1$, reason why we choose it. 

During training, we updated iteratively the weights of the whole model in pursuance of obtaining the minimum focal loss \cite{lin2017focal}. We evaluated the network at the end of each epoch using the validation dataset. The model which achieve the best accuracy in the validation dataset was retained to be evaluated with the testing dataset. The maximum number of epochs was set equal to $500$. 

Since the SHHS database records were obtained in subjects during a polysomnography, with the aim of assess the sleep quality, in most registers there is a class imbalance. Specifically,  most of the total segments in the recording correspond to sleep stage. The relationship between the sleep and awake time was showed in Table \ref{tab:shhs1}. There are many methods to prevent the classifier from biasing towards the majority class. In this work, we used the focal loss approach, which proposes a modification of the cross entropy loss, focusing the training on hard-examples. We varied the parameter $\gamma$ from $0$ to $10$ in unitary steps, which down-weight the contribution of the easy examples. Further, we varied the $\alpha$ parameter in range $[0, \, 1]$, which increases the weight of the minority class in the total loss function. The best performance was obtained with $\alpha=0.4$ and $\gamma=5$. A complete description of focal loss can be found in \cite{lin2017focal}.

\subsection{Designed and baseline networks}

For the design of the architecture, we varied methodically both the parameters of the representation learning and sequence modeling parts. For the representation learning part, we varied the whole residual blocks parameters, namely the size filter, the number of output channels, the stride and the dilation of the convolutions. Furthermore, we tried with different number of residual blocks. The best performance was obtained with the architecture shown in Fig. \ref{fig1}. For the sequence modeling part, we varied the number of heads $h$ in the multi-head attention, the number of encoder layers, and the dimension in the FNN of the the transformer. The parameter combination to achieve the best performance is shown in Fig. \ref{fig2}. 

Since we used a transformer-based design for sequence modeling, the number of possible architectures to address the problem is rather limited. Thus, we can vary the parameters of the transformer that we have already modified, but the architecture contains a repeating basic block. Nevertheless, the possibility of architectures for the representation learning part is considerable. In order to assess the contribution of this part to the whole network, we developed two baselines architectures based in different methods for comparison.

First, we evaluated a single-layer FNN, which receives as input $30$-consecutive samples of the sequence and maps them to a $d_\textrm{model}$, using hyperbolic tangent as activation function.  Through this simple configuration we evaluated the contribution of the sequence modeling part. Furthermore, this experiment reveals the need for more elaborate architectures to allow the feature extraction in the representation learning part, which leads us to the second group of baselines based on convolutional neural network.

We tried with several configurations based on convolutional neural networks, the layers of which sequentially performs the following operations: 1D convolution, dropout, batch normalization and \textrm{ReLU} activation. We modified the number of considered layers, the filter size and stride parameters of the convolution. The best accuracy with these types of networks was achieved with an analogous architecture to that presented in Fig. \ref{fig1}, namely four layers with kernel size equal to $3$ and strides equal to $5$, $3$, $2$ and $1$, respectively. Also, the dimensionality increase from $1$ to $d_\textrm{model}$ was obtained in the same manner that for the TCN architecture.

\section{Results} \label{sec:results}

\begin{table*}[t!]
\scriptsize
\centering
\caption{Performance of the networks in train, validation and test datasets.}
\begin{tabular}{c c c c c c c c c c c c }
\hline 
  Method                          & Signal   & Subjects & Acc	 & Se     & Sp     & PPV    & NPV  & $\kappa$  & Training time (min) \\
  \hline 	
  TCN + Transformer               & HR (PPG) & $5000$   & $90.0$ & $94.9$ & $78.1$ & $91.3$ & $85.2$ & $0.73$ & $558.9$ \\
  CNN + Transformer               & HR (PPG) & $5000$   & $86.5$ & $95.0$ & $65.3$ & $87.3$ & $83.8$ & $0.62$ & $304.4$ \\
  FNN + Transformer               & HR (PPG) & $5000$   & $84.8$ & $94.5$ & $63.1$ & $85.8$ & $80.7$ & $0.59$ & $25.9$  \\
  
                &  &    &  &  &  &   &  &  &   \\
                
  CWT feat. + RF \cite{fraiwan2012automated}&EEG   &$16$      & $93.4$ & $89.6$ & $95.9$ & $93.3$ & $93.5$ & $0.86$ & -- \\
  
  CNN + RNN \cite{supratak2017deepsleepnet}& EEG & $20$ & $94.3$ & $83.4$ & $96.8$ & $86.0$ & $96.2$ & $0.81$ & -- \\
  
                &  &    &  &  &  &   &  &  &   \\
                
  Feat. + SVM + RFE \cite{adnane2012sleep} &ECG     & $18$      & $80.0$ & $84.5$ & $69.1$ & $87.0$ & $64.5$ & $0.52$ & -- \\

  CNN \cite{malik2018sleep}       & ECG     & $27$      & $83.1$ & $89.4$ & $52.4$ & $90.1$ & $50.5$ & $0.41$ & -- \\
  
              &  &    &  &  &  &   &  &  &   \\
              
  Feat. + SVM \cite{uccar2016automatic}   &HRV      & $10$     & $73.1$ & $74.0$ & $72.0$ & --     & -- & $0.46$ & -- \\
  Feat. + SVM \cite{uccar2016automatic}   &PPG      & $10$     & $76.8$ & $76.0$ & $77.0$ & --     & -- & $0.53$ & -- \\
  Feat. + kNN \cite{uccar2016automatic}   &PPG + HRV& $10$     & $79.4$ & $77.0$ & $81.0$ & --	  & -- & $0.59$ & -- \\

  CNN \cite{malik2018sleep}       & PPG     & $27$      & $84.2$ & $90.9$ & $51.5$  & $90.1$ & $53.6$ & $0.43$ & -- \\
    
  2 GRU-(128) \cite{casal2019rnn} & HR (PPG) & $5000$   & $89.1$ & $94.2$ & $76.8$ & $90.7$ & $83.6$ & $0.71$ & $8821.7$\\
  Feat. + FFS-SVM   \cite{casal2019hely}  & HR (PPG) & $5000$   & $73.7$ & $80.9$ & $54.6$ & $84.8$ & $0.49$ & $0.44$     & --\\
  2 GRU-(256) \cite{casal2019rnn}&HR + SpO$_2$ (PPG)&$5000$&$90.1$&$94.1$ & $80.3$ & $92.1$ & $84.7$ & $0.74$ & $12376.7$ \\     

                &  &    &  &  &  &   &  &  &   \\
  
  Feat. + RNN \cite{fonseca2020automatic}& ECG + acc  & $778$ 	& $90.1$ & $94.0$ & $72.9$ & --     & $75.1$ & $0.65$  & -- \\   

  Feat. + LDA \cite{beattie2017estimation}& PPG + acc & $60$ 	& $90.6$ & $94.6$ & $69.3$ & $94.3$ & $70.5$ & $0.64$ & -- \\
\hline
\end{tabular}
\label{tab:Performance_measures}
\end{table*}

In this section we show the obtained results and a detailed comparison with the designed baselines as well as published works in the area. The performance of the models was evaluated using accuracy, sensitivity, specificity, positive predictive value, negative predictive value, and Cohen's Kappa coefficient $\kappa$ \cite{sokolova2009systematic}. All these measures were selected with the aim of presenting metrics which are not affected by the imbalance between awake/sleep classes, this means the skew class does not modify the value \cite{fawcett2006introduction}.  First, we calculated all these metrics for each patient. Then, we obtained an average across patient for every metric. In this work he awake stage was consider as positive class. 

\begin{figure*}[t!]
\centering
\includegraphics[width=0.95\textwidth]{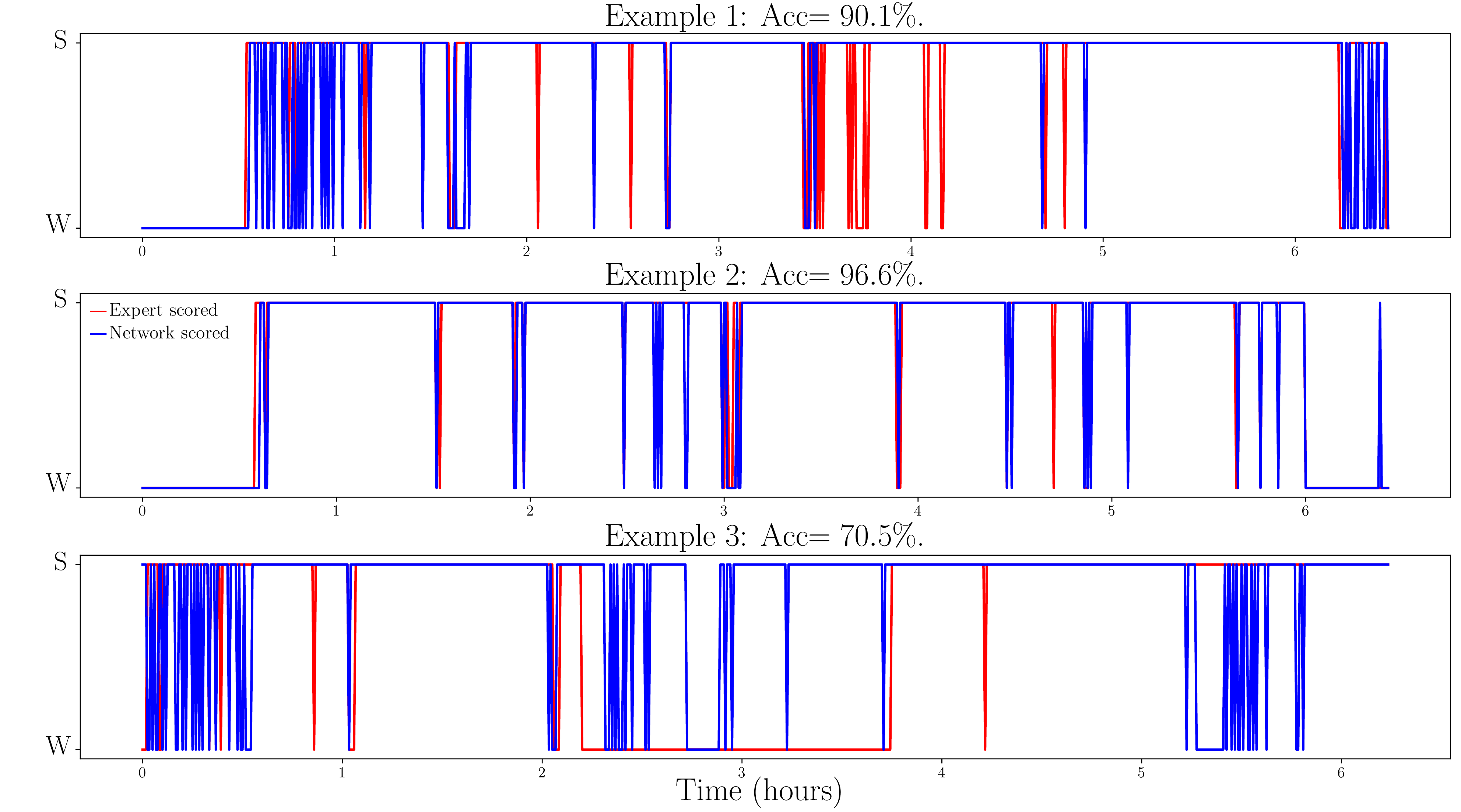}
\caption{Three example of obtained-simplified hypnograms using the developed network. The top-graph shows an average-performance example, the middle-graph shows an high-performance example and the bottom-graph shows an low-performance example. The red and blue lines are the expert scoring and network scoring, respectively.}
\label{fig:Hipnogram_transformer}
\end{figure*}

We evaluated the performance of our methods using the test dataset. In Table \ref{tab:Performance_measures}, it is shown an extensive comparison between the best configuration of the here developed method and baselines, our two previous works and other published works in the area. Along with the performance metrics, we report the used signal, the number of patients in the database, and total training time.

It is worth clarifying that several previous works presented in Table \ref{tab:Performance_measures} discriminated sleep stages in detail, according with the sleep staging guidelines. For comparative purposes, the stages N1, N2, N3 and REM were grouped into S stage. Furthermore, some metrics were deduced from the results showed by the authors when they were not presented explicitly. 

The performance is improved by using TCN instead of CNN or FNN. On the other hand, the total training time of a more complex network is greater. However, it is observed that the training time of the here-developed method is an order of magnitude lower than in RNN for the same application. Further, the former were trained for $500$ epochs to reach the optimum, whereas the later in \cite{casal2019rnn} only need $100$. The difference is due to the usage of L2 penalty in the present work, which leads to a smoother convergence but demands a longer training time.

Figure \ref{fig:Hipnogram_transformer} shows three obtained simplified-hypnograms using the ``best'' network. The first hypnogram is an average-performance example, the second one is an high-performance example and the third one is low-performance example. 

\begin{figure}[!t]
\centering
\includegraphics[width=0.65\textwidth]{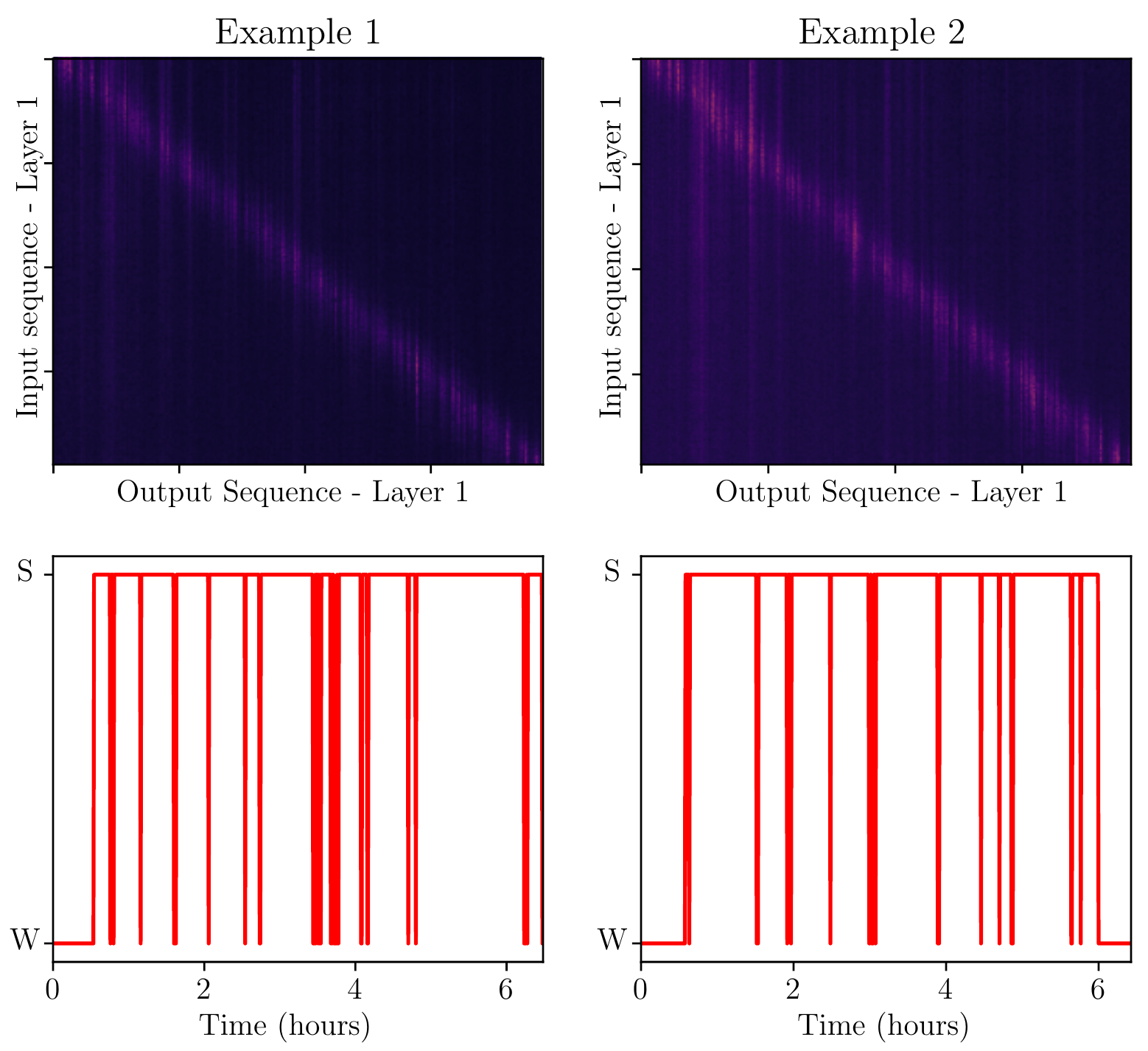}
\caption{Attention maps of two different subjects. The average attention maps for layer 1 is shown, as well as the simplified hypnogram.}
\label{fig:heatmaps}
\end{figure}

Finally, since we have used an architecture based on attention mechanisms, we can obtain an interpretation about which input epochs are important to score the sleep stage by observing the attention maps. The attention maps --also called heatmaps-- are obtained by graphing the matrix resulting from solving the softmax in the equation \ref{eq:attention}. This matrix has values between 0 and 1, which quantify the ``attention'' in every input epoch for each output scores. Since the optimal architecture found in this work consists of two layers of transformers with $8$ multihead-attention mechanisms, there are $8$ attention matrices for each layer. For the sake of simplicity and interpretability, Fig. \ref{fig:heatmaps} shows a single matrix for each layer, which was calculated as the average of the $8$ multi-head attention matrices. That is, the overall attention of the layer is shown. We can see the network uses approximately $100$ input segments to determine the current stage (on the diagonal of each figure). In the figure it can also be seen that in transition moments --from W to S and vice versa-- it is normal for the network to increase attention to a larger number of input segments, which allows us to think that it needs a greater receptive field. Fig. \ref{fig:heatmaps} shows the attention matrices for two different subjects.  

In the next section we will address an exhaustive discussion about sleep staging and a comprehensive comparison with other methods.

\section{Discussion} \label{sec:discussion}

We proposed a neural network architecture based on TCN and transformers to classify the sleep stage in awake and sleep from raw HR signals provided by a pulse oximeter. Although the obtained performance is comparable with the state of the art works which have used the same signals, the training time is considerably faster. This last statement along with the solely use of a pulse oximeter make this network potentially useful to be part of home-based and wearable devices for sleep staging.

The first algorithms which attempt to automatically classify sleep are emerged as a natural approach to the guidelines for scoring sleep stages, trying to facilitate and speed up the visual scoring. Such methods can be grouped into two categories, namely multi-channel and single-channel algorithms. Multi-channel approaches are based on one or more EEG channels, EMG and/or EOG, obtaining better performances than the ones based on single-channel EEG. Nevertheless, these are inadequate for home sleep testing due to complexity. In addition, they can affect the patient falling asleep \cite{boostani2017comparative}.

We include in the comparison two previous work in which EEG is used. Fraiwan et al. \cite{fraiwan2012automated} developed a method to classify sleep stages according with AASM rules using time-frequency distribution and random forest as a classifier. This work used hand-crafted features. They obtained the best performance in a comprehensive review published in 2017 \cite{boostani2017comparative}. Then, Supratak et al. \cite{supratak2017deepsleepnet} proposed a neural network that use CNN to automatically learn features from a raw single-channel of EEG, and RNN to classify according to AASM guidelines. Both results are the gold standard for sleep staging classification. The obtained performances should be our final reference to evaluate the effectiveness of our proposed method.

In addition to the automatic sleep staging based on a single-channel of EEG, algorithms based on cardiac-related signals have been developed in order to simplify the home-based screening. These works exploit the information about sleep staging related with the HR regulation \cite{penzel2003dynamics}. Adnane et. al \cite{adnane2012sleep} developed a classic machine learning method using ECG to classify the sleep stage in awake or asleep. The method is based on hand-craft feature extraction, followed by a support vector machine classifier with recursive feature elimination. This work designed a specific-subject scheme. It means that a part of the subject data was used to train the classifier and the remaining part was used to test it. In general, the performance decreases when using different subjects for training and testing. Malik et al. \cite{malik2018sleep} designed a CNN classify wake/sleep status using the instantaneous heart rate (IHR) obtained from ECG. This work used automatic learned features. Furthermore, the network also was tested using PPG to calculate IHR. The authors of this work trained and tested the network on different dataset, assessing the proficiency transfer among different devices and dataset. The obtained performances were similar. For the sake of readability, we only reported one of the ECG and PPG results for the same dataset. 

The pulse oximeter is a simpler device than the ECG from which cardiac information can also be inferred. U{\c{c}}ar et al. \cite{uccar2016automatic} developed a classic method that use PPG, HR variability and a combination of both to classify in asleep and awake using SVM or k-nearest neighbour. We showed the best result of each combination of inputs. To continue in the same order in which we have presented the previous works, Malik et al. \cite{malik2018sleep} applied their method designed for ECG using PPG signals, which was already discussed in the previous paragraph.

Beattie et al. \cite{beattie2017estimation} combined PPG signals with accelerometer to classify the sleep stages in W, light sleep, deep sleep and REM sleep. By adding the accelerometer signals, they have additional information, but this sensors can be embedded whitout increasing the device complexity. Thus, the algorithm is still useful for portable systems and wereables. Although this work used hand-craft features, it achieves a remarkable performance, which is comparable to ours. 

Finally, Fonseca et al. \cite{fonseca2020automatic} used ECG and accelerometer signals to classify the sleep stages according with AASM guidelines. The authors used hand-craft features combined with a recurrent neural network architecture, obtaining significant results. Furthermore, this work trained and tested the developed network in different databases. 

The comparison between different algorithms is a complicated process because they are designed based on different datasets and type of signals. In spite of this, the obtained results are comparable with the state of the art, excepting those which used EEG. It is worth noting that we used a larger database and simpler signals than those used in other works. 

One of the remarkable aspects of our proposed method is the possibility to obtain an interpretation of the results. As could be seen in the Fig. \ref{fig:heatmaps}, the first layer of the transformer uses approximately $50$ previous and $50$ after epochs to score a sleep stage (the height on the diagonal). If we consider that the epochs last $30$ seconds, this would correspond to a $50$-minutes ``attention window'' in total. However, it must be taken into account that the input to the transformer is the representation obtained by the first part of the network, therefore each epoch is influenced by some neighboring epochs due to the dilated convolutions. In spite of this, we can realize the receptive-field size that influence the decision. Also, this result contributes to the discussion started in \cite{casal2019hely} about the length of the segment to be classified. Moreover, it can be seen how this ``attention window''  grows in the transition moments between W and S, meaning that the network needs to take into account more segments of the sequence to determine these changes. Finally, it remains to mention that the interpretation is more complex for the second-transformer layer of transformer, generally presenting a vertical structure in the attention maps. This could lead to think that the network is oversized, and with a single-transformer layer it would be able to obtain the same result. Nevertheless, as mentioned in the performed experiments, the here presented architecture is the one that obtained the best results for all the parameters explored. On the other hand, it can be seen from this that a larger network would not be necessary. Here, several factors that can modify the convergence of algorithms must be considerated, such as optimization algorithms, regularization and normalization methods. This may be one of the reasons why a possibly oversized network has a better performance than the smaller network, as is discussed in \cite{he2016deep, wu2019wider}.

\section{Conclusions}
We proposed a deep learning model based on attention mechanisms to classify sequence to sequence the sleep stages in awake or asleep using HR signals provided by a commercial pulse oximeter. Our model is based on TCN to learn features from HR, and transformers to model the sequence, allowing the classification. The network was trained end-to-end using a scheme of train, validation and test dataset obtained from the SHHS database. The performed experiments with designed baselines showed that the TCN architecture outperforms the CNN or FNN architectures to extract automatically features. The results achieved by our network are comparable with the state-of-the-art performance using simpler signals and a bigger database. Also, the training time was significantly reduced, especially compared with RNN-based methods. The here developed work constitutes an advance for the simplified diagnosis of sleep disorders. As a future work, we will combine this work with methods to detect apnea/hypopnea events based on pulse oximeter with the aim of improving the state-of-the-art results in screening of sleep obstructive apnea/hypopnea syndrome.
%
%
%

\bibliographystyle{elsarticle-num}
\bibliography{refs}

\begin{thebibliography}{10}
\expandafter\ifx\csname url\endcsname\relax
  \def\url#1{\texttt{#1}}\fi
\expandafter\ifx\csname urlprefix\endcsname\relax\def\urlprefix{URL }\fi
\expandafter\ifx\csname href\endcsname\relax
  \def\href#1#2{#2} \def\path#1{#1}\fi

\bibitem{stickgold2005sleep}
R.~Stickgold, Sleep-dependent memory consolidation, Nature 437~(7063) (2005)
  1272--1278.

\bibitem{sateia2014international}
M.~J. Sateia, International classification of sleep disorders: highlights and
  modifications, Chest Journal 146~(5) (2014) 1387--1394.

\bibitem{skaer2010economic}
T.~L. Skaer, D.~A. Sclar, Economic implications of sleep disorders,
  Pharmacoeconomics 28~(11) (2010) 1015--1023.

\bibitem{rechtschaffen1968manual}
A.~Rechtschaffen, A manual of standardized terminology, technique and scoring
  system for sleep stages of human subjects, Public Health Service (1968).

\bibitem{berry2012aasm}
R.~B. Berry, R.~Brooks, C.~E. Gamaldo, S.~M. Harding, C.~Marcus, B.~Vaughn, The
  {AASM} manual for the scoring of sleep and associated events, Rules,
  Terminology and Technical Specifications, Darien, Illinois, American Academy
  of Sleep Medicine (2012).

\bibitem{norman2000interobserver}
R.~G. Norman, I.~Pal, C.~Stewart, J.~A. Walsleben, D.~M. Rapoport,
  Interobserver agreement among sleep scorers from different centers in a large
  dataset., Sleep 23~(7) (2000) 901--908.

\bibitem{supratak2017deepsleepnet}
A.~Supratak, H.~Dong, C.~Wu, Y.~Guo, Deepsleepnet: a model for automatic sleep
  stage scoring based on raw single-channel eeg, IEEE Transactions on Neural
  Systems and Rehabilitation Engineering 25~(11) (2017) 1998--2008.

\bibitem{phan2019seqsleepnet}
H.~Phan, F.~Andreotti, N.~Cooray, O.~Y. Ch{\'e}n, M.~De~Vos, Seqsleepnet:
  end-to-end hierarchical recurrent neural network for sequence-to-sequence
  automatic sleep staging, IEEE Transactions on Neural Systems and
  Rehabilitation Engineering 27~(3) (2019) 400--410.

\bibitem{fraiwan2012automated}
L.~Fraiwan, K.~Lweesy, N.~Khasawneh, H.~Wenz, H.~Dickhaus, Automated sleep
  stage identification system based on time--frequency analysis of a single eeg
  channel and random forest classifier, Computer methods and programs in
  biomedicine 108~(1) (2012) 10--19.

\bibitem{fuhrman2012symptoms}
C.~Fuhrman, B.~Fleury, X.-L. Nguy{\^e}n, M.-C. Delmas, Symptoms of sleep apnea
  syndrome: high prevalence and underdiagnosis in the french population, Sleep
  medicine 13~(7) (2012) 852--858.

\bibitem{penzel2003dynamics}
T.~Penzel, J.~W. Kantelhardt, L.~Chung-Chang, K.~Voigt, C.~Vogelmeier, Dynamics
  of heart rate and sleep stages in normals and patients with sleep apnea,
  Neuropsychopharmacology 28~(S1) (2003) S48.

\bibitem{aeschbacher2016heart}
S.~Aeschbacher, M.~Bossard, T.~Schoen, D.~Schmidlin, C.~Muff, A.~Maseli, J.~D.
  Leuppi, D.~Miedinger, N.~M. Probst-Hensch, A.~Schmidt-Trucks{\"a}ss, et~al.,
  Heart rate variability and sleep-related breathing disorders in the general
  population, The American Journal of Cardiology 118~(6) (2016) 912--917.

\bibitem{adnane2012sleep}
M.~Adnane, Z.~Jiang, Z.~Yan, Sleep--wake stages classification and sleep
  efficiency estimation using single-lead electrocardiogram, Expert Systems
  with Applications 39~(1) (2012) 1401--1413.

\bibitem{yucelbacs2018automatic}
{\c{S}}.~Y{\"u}celba{\c{s}}, C.~Y{\"u}celba{\c{s}}, G.~Tezel,
  S.~{\"O}z{\c{s}}en, {\c{S}}.~Yosunkaya, Automatic sleep staging based on
  {SVD}, {VMD}, {HHT} and morphological features of single-lead {ECG} signal,
  Expert Systems with Applications 102 (2018) 193--206.

\bibitem{malik2018sleep}
J.~Malik, Y.-L. Lo, H.-t. Wu, Sleep-wake classification via quantifying heart
  rate variability by convolutional neural network, Physiological measurement
  39~(8) (2018) 085004.

\bibitem{uccar2016automatic}
M.~K. U{\c{c}}ar, M.~R. Bozkurt, C.~Bilgin, K.~Polat, Automatic sleep staging
  in obstructive sleep apnea patients using photoplethysmography, heart rate
  variability signal and machine learning techniques, Neural Computing and
  Applications (2016) 1--16.

\bibitem{casal2019hely}
R.~Casal, L.~E. Di~Persia, G.~Schlotthauer, Sleep-wake stages classification
  using heart rate signals from pulse oximetry, Heliyon 5~(10) (2019) e02529.

\bibitem{beattie2017estimation}
Z.~Beattie, Y.~Oyang, A.~Statan, A.~Ghoreyshi, A.~Pantelopoulos, A.~Russell,
  C.~Heneghan, Estimation of sleep stages in a healthy adult population from
  optical plethysmography and accelerometer signals, Physiological Measurement
  38~(11) (2017) 1968.

\bibitem{casal2019rnn}
R.~Casal, L.~E. Di~Persia, G.~Schlotthauer, Classifiying sleep-wake stages
  through recurrent neural networks using pulse oximetry signals, Paper Nuevo
  X~(xx) (2020) zzzx--zzzy.

\bibitem{lecun2015deep}
Y.~LeCun, Y.~Bengio, G.~Hinton, Deep learning, Nature 521~(7553) (2015) 436.

\bibitem{vaswani2017attention}
A.~Vaswani, N.~Shazeer, N.~Parmar, J.~Uszkoreit, L.~Jones, A.~N. Gomez,
  {\L}.~Kaiser, I.~Polosukhin, Attention is all you need, in: Advances in
  neural information processing systems, 2017, pp. 5998--6008.

\bibitem{bai2018empirical}
S.~Bai, J.~Z. Kolter, V.~Koltun, An empirical evaluation of generic
  convolutional and recurrent networks for sequence modeling, arXiv preprint
  arXiv:1803.01271 (2018).

\bibitem{nair2010rectified}
V.~Nair, G.~E. Hinton, Rectified linear units improve restricted boltzmann
  machines, in: Proceedings of the 27th international conference on machine
  learning (ICML-10), 2010, pp. 807--814.

\bibitem{he2016deep}
K.~He, X.~Zhang, S.~Ren, J.~Sun, Deep residual learning for image recognition,
  in: Proceedings of the IEEE conference on computer vision and pattern
  recognition, 2016, pp. 770--778.

\bibitem{springenberg2014striving}
J.~T. Springenberg, A.~Dosovitskiy, T.~Brox, M.~Riedmiller, Striving for
  simplicity: The all convolutional net, arXiv preprint arXiv:1412.6806 (2014).

\bibitem{ioffe2015batch}
S.~Ioffe, C.~Szegedy, Batch normalization: Accelerating deep network training
  by reducing internal covariate shift, arXiv preprint arXiv:1502.03167 (2015).

\bibitem{hinton2012improving}
G.~E. Hinton, N.~Srivastava, A.~Krizhevsky, I.~Sutskever, R.~R. Salakhutdinov,
  Improving neural networks by preventing co-adaptation of feature detectors,
  arXiv preprint arXiv:1207.0580 (2012).

\bibitem{ba2016layer}
J.~L. Ba, J.~R. Kiros, G.~E. Hinton, Layer normalization, arXiv preprint
  arXiv:1607.06450 (2016).

\bibitem{redline1998methods}
S.~Redline, M.~H. Sanders, B.~K. Lind, S.~F. Quan, C.~Iber, D.~J. Gottlieb,
  W.~H. Bonekat, D.~M. Rapoport, P.~L. Smith, J.~P. Kiley, et~al., Methods for
  obtaining and analyzing unattended polysomnography data for a multicenter
  study, Sleep 21~(7) (1998) 759--768.

\bibitem{nieto1997sleep}
E.~J. Nieto, G.~T. O'Connor, D.~M. Rapoport, S.~Redline, The sleep heart health
  study: design, rationale, and methods, Sleep 20~(12) (1997) 1077--1085.

\bibitem{kingma2014adam}
D.~P. Kingma, J.~Ba, Adam: A method for stochastic optimization, arXiv preprint
  arXiv:1412.6980 (2014).

\bibitem{srivastava2014dropout}
N.~Srivastava, G.~Hinton, A.~Krizhevsky, I.~Sutskever, R.~Salakhutdinov,
  Dropout: a simple way to prevent neural networks from overfitting, The
  journal of machine learning research 15~(1) (2014) 1929--1958.

\bibitem{lin2017focal}
T.-Y. Lin, P.~Goyal, R.~Girshick, K.~He, P.~Doll{\'a}r, Focal loss for dense
  object detection, in: Proceedings of the IEEE international conference on
  computer vision, 2017, pp. 2980--2988.

\bibitem{fonseca2020automatic}
P.~Fonseca, M.~M. van Gilst, M.~Radha, M.~Ross, A.~Moreau, A.~Cerny,
  P.~Anderer, X.~Long, J.~P. van Dijk, S.~Overeem, Automatic sleep staging
  using heart rate variability, body movements, and recurrent neural networks
  in a sleep disordered population, Sleep (2020).

\bibitem{sokolova2009systematic}
M.~Sokolova, G.~Lapalme, A systematic analysis of performance measures for
  classification tasks, Information Processing \&amp; Management 45~(4) (2009)
  427--437.

\bibitem{fawcett2006introduction}
T.~Fawcett, An introduction to {ROC} analysis, Pattern recognition letters
  27~(8) (2006) 861--874.

\bibitem{boostani2017comparative}
R.~Boostani, F.~Karimzadeh, M.~Nami, A comparative review on sleep stage
  classification methods in patients and healthy individuals, Computer methods
  and programs in biomedicine 140 (2017) 77--91.

\bibitem{wu2019wider}
Z.~Wu, C.~Shen, A.~Van Den~Hengel, Wider or deeper: Revisiting the resnet model
  for visual recognition, Pattern Recognition 90 (2019) 119--133.

\end{thebibliography}

\end{document}